\documentclass[aoas,MSNbibl,nameyear,seceqn,dvips]{arximspdf}
\usepackage{dcolumn}

\doi{10.1214/12-AOAS555} 
\volume{6}
\issue{4}
\pubyear{2012}
\firstpage{1814}
\lastpage{1837}

\makeatletter
\newcolumntype{d}[1]{D{.}{.}{#1}}
\newcommand{\eqref}[1]{(\ref{#1})}
\makeatother

\begin{document}
\begin{frontmatter}

\title{Addressing missing data mechanism uncertainty using multiple-model multiple imputation: Application to a longitudinal clinical trial}
\runtitle{Multiple-model multiple imputation}

\begin{aug}
\author[A]{\fnms{Juned} \snm{Siddique}\corref{}\thanksref{t1}\ead[label=e1]{siddique@northwestern.edu}},
\author[B]{\fnms{Ofer} \snm{Harel}\thanksref{t2}\ead[label=e2]{oharel@stat.uconn.edu}}
\and
\author[C]{\fnms{Catherine M.} \snm{Crespi}\thanksref{t3}\ead[label=e3]{ccrespi@ucla.edu}}
\thankstext{t1}{Supported in part by AHRQ Grant R03 HS018815 and NIH
Grants R01 MH040859 and K07 CA154862.}
\thankstext{t2}{Supported in part by NIH Grant K01 MH087219.}
\thankstext{t3}{Supported in part by NIH Grants CA133081 and UL1TR000124.}
\runauthor{J. Siddique, O. Harel and C. M. Crespi}
\affiliation{Northwestern University, University of Connecticut and
University~of~California Los Angeles}
\address[A]{J. Siddique\\
Department of Preventive Medicine\\
Northwestern University\\
Feinberg School of Medicine\\
Chicago, Illinois 60611\\
USA\\
\printead{e1}}

\address[B]{O. Harel\\
Department of Statistics\\
University of Connecticut\\
Storrs, Connecticut 06269\\
USA\\
\printead{e2}}

\address[C]{C. M. Crespi\\
Department of Biostatistics\\
University of California Los Angeles\\
School of Public Health\\
Los Angeles, California 90095\\
USA\\
\printead{e3}}

\end{aug}

\received{\smonth{4} \syear{2011}}
\revised{\smonth{1} \syear{2012}}

%
\begin{abstract}
We present a framework for generating multiple imputations for
continuous data when the missing data mechanism is unknown. Imputations
are generated from more than one imputation model in order to
incorporate uncertainty regarding the missing data mechanism. Parameter
estimates based on the different imputation models are combined using
rules for nested multiple imputation. Through the use of simulation, we
investigate the impact of missing data mechanism uncertainty on
post-imputation inferences and show that incorporating this uncertainty
can increase the coverage of parameter estimates. We apply our method
to a longitudinal clinical trial of low-income women with depression
where nonignorably missing data were a concern. We show that different
assumptions regarding the missing data mechanism can have a substantial
impact on inferences. Our method provides a simple approach for
formalizing subjective notions regarding nonresponse so that they can
be easily stated, communicated and compared.
\end{abstract}

%
\begin{keyword}
\kwd{Nonignorable}
\kwd{NMAR}
\kwd{MNAR}
\kwd{not missing at random}
\kwd{missing not at random}.
\end{keyword}

\end{frontmatter}
%

\section{Introduction}
The longitudinal clinical trial is a powerful design for estimating and
comparing rates of change over time in two or more treatment groups.
However, measuring participants repeatedly over time provides repeated
opportunities for participants to miss measurement occasions. Missing
values are a problem in most longitudinal studies and a variety of
methods have been developed to produce valid inferences in the presence
of missing data. In particular, multiple imputation---where missing
values are replaced with two or more plausible values---has gained
widespread acceptance in recent years and is a common and flexible
approach for handling missing data.

When dealing with missing data, special concern must be given to the
process that gave rise to the missing data, referred to as the missing
data mechanism. Most methods for generating multiple imputations, both
fully-parametric methods [\citet{liu95}, \citet{schafer97}] and semi-parametric
methods [\citet{raghunathan01}, \citet{schenker96}, \citet{siddique08a}, \citet{vanbuuren07}],
assume the missing data mechanism is \textit{ignorable} as described
by \citet{rubin76}, where the probability that a value is missing does
not depend on unobserved information such as the value itself. When
data are nonignorably missing, that is, the probability that a value is
missing \textit{does} depend on unobserved information, the model for
generating imputations must take into account the missing data
mechanism. The role of nonignorability assumptions has been discussed
in the context of a variety of applied settings; see, for
example, \citeauthor{little02} [(\citeyear{little02}), chapter~15],
\citet{belin93}, \citet{rubin95}, \citet{schafer02}, \citet{wachter93} and \citet{demirtas03}.

Nonignorably missing data is of particular concern in depression trials
because it is very likely that the reason for a participant missing an
assessment or dropping out of a study is related to their underlying
depression status [\citet{blackburn81}, \citet{elkin89}, \citet{warden09}]. For example, a
depressed participant may feel like the intervention is not working for
them and may be unwilling to sit through an interview and/or answer the
phone. Conversely, a high-functioning, nondepressed participant may
feel like he no longer needs to remain in the trial or may not be
available for an assessment because he is busy working, shopping or
socializing. Failure to take into account the missing data mechanism
may result in inferences that make a treatment appear more or less
effective. Failure to incorporate uncertainty regarding the missing
data mechanism may result in inferences that are overly precise given
the amount of available information [\citet{demirtas03}].

Since a nonignorable missing data mechanism depends on unobserved data,
there is little information available to correctly model this process.
A~common approach in such cases is to perform a sensitivity analysis,
drawing inferences based on a variety of assumptions regarding the
missing data mechanism [\citet{daniels08}]. There is a broad literature
on sensitivity analyses for exploring unverifiable missing data
assumptions [see \citet{ibrahim09} and discussion for a review]. One
approach begins with the specification of a full-data distribution,
followed by examination of inferences across a range of values for one
or more unidentified parameters [\citet{daniels08}, \citet{molenberghs01}, \citet{rubin77}, \citet{scharfstein99},
\citet{vansteelandt06}].

When a decision is required, a drawback of sensitivity analysis is that
it produces a range of answers rather than a single answer [\citet
{scharfstein99}]. Several authors have proposed model-based methods for
obtaining a final inference. This approach involves placing an
informative prior distribution on the unidentified parameters that
characterize assumptions about the missing data mechanism. Then,
inferences are drawn that incorporate a range of assumptions regarding
the missing data mechanism [\citet{daniels08}, \citet{forster98}, \citet{kaciroti06}, \citet{rubin77}].

An alternative approach for handling data with nonignorable missingness
is multiple imputation. Multiple imputation methods have several
advantages over model-based methods for analyzing data with missing
values: they allow for standard complete-data methods of analysis to be
performed once the data have been imputed [\citet{little02}], and
auxiliary variables that are not part of the analysis procedure can be
incorporated into the imputation procedure to increase efficiency and
reduce bias [\citet{collins01}].

Methods for multiple imputation with nonignorably missing data include
those of \citet{carpenter07} who use a reweighting approach to
investigate the influence of departures from the ignorable assumption
on parameter estimates. \citet{vanbuuren99} perform a sensitivity
analysis with multiply imputed data using offsets to explore how robust
their inferences are to violations of the assumption of ignorability. A
limitation of these approaches is that they do not take into account
uncertainty regarding the missing data mechanism. Instead, they provide
a range of inferences for various ignorability assumptions.

\citet{landrum01} develop an imputation procedure that allows for model
uncertainty to be reflected in the multiple imputations for those cases
in which no one imputation model is clearly the best model by drawing
imputations from more than one model. However, their procedure assumes
ignorably missing data. \citet{siddique08b} use a nonignorable
approximate Bayesian bootstrap to generate multiple imputations
assuming nonignorability. Each set of imputations is based on a
different assumption regarding the missing data mechanism in order to
incorporate missing data mechanism uncertainty. However, \citet
{siddique08b} use conventional multiple imputation combining rules
which are not appropriate when imputations are generated from different
posterior distributions because they do not take into account the
additional uncertainty due to using more than one imputation model.

In this paper we describe a new multiple imputation approach for
estimating parameters and their associated confidence intervals in the
presence of nonignorable nonresponse. Our goal is to develop a multiple
imputation framework analogous to model-based methods such as those
of \citet{rubin77}, \citet{forster98}\vadjust{\goodbreak} and \citet{daniels08} that
incorporate a range of ignorability assumptions into one inference.
Rather than attempting the hopeless objective of correctly modeling the
missing data mechanism, we generate our imputations using multiple
imputation models and then use specialized combining rules to generate
inferences that incorporate missing data mechanism uncertainty.
Imputations are generated in three steps: (1)~a~distribution of models
incorporating ignorable and/or nonignorable mechanisms is specified;
(2) a model is drawn from this distribution; (3)~multiple imputations
are generated from the model selected in Step 2. Steps 2 and 3 are then
repeated, thereby generating multiple-model multiple imputations. The
nested imputation combining rules of \citet{shen00} are used to combine
inferences across multiple imputations so that between-model
uncertainty is incorporated into the standard errors of parameter
estimates.\looseness=-1

The outline for the rest of this paper is as follows. In Section \ref
{S:wecare} we describe the WECare study, a longitudinal depression
treatment trial that motivated this work. In Section \ref{S:methods} we
describe methods for generating multiple-model multiple imputations for
continuous data in order to incorporate missing data mechanism
uncertainty and describe the nested imputation combining the rules
of \citet{shen00}. In addition, we develop a method of quantifying the
contribution of missing data mechanism uncertainty to the overall rate
of missing information. Section \ref{S:sim} describes the design of a
simulation study and Section \ref{S:sim_results} presents the results
of the simulation study. In Section \ref{S:apply} we apply our approach
to the WECare study. Section~\ref{S:discuss} provides a discussion.

Closely related to the concept of ignorability are the missing data
mechanism taxonomies ``missing at random'' (MAR) and ``not missing at
random'' (NMAR). MAR requires that the probability of missingness
depends on observed values only, while ignorability includes the
additional assumption that the parameters that generate the data and
the parameters governing the missing data mechanism are distinct [\citet{little02},
\citet{rubin76}]. While distinctness of these two sets of parameters
cannot always be assumed (particularly in time to event data), for the
purposes of this paper we will use the terms MAR and ignorable
interchangeably and the terms NMAR and nonignorable interchangeably.

\section{Motivating example: The WECare study}\label{S:wecare}
The Women Entering Care (WECare) Study investigated depression outcomes
during a 12-month period in which 267 low-income mostly minority women
in the suburban Washington, DC area were treated for depression. The
participants were randomly assigned to one of three treatment groups:
Medication, Cognitive Behavioral Therapy (CBT) or treatment-as-usual
(TAU), which consisted of referral to a community provider. Depression
was measured every month through a phone interview using the Hamilton
Depression Rating Scale (HDRS).

Information on ethnicity, income, number of children, insurance and
education was collected during the screening and the baseline
interviews. All screening and baseline data were complete except for
income, with 10 participants missing data on income. After baseline,
the percentage of missing interviews ranged between 24\% and 38\%
across months.

Outcomes for the first six months of the study were reported in \citet
{miranda03}. In that paper the primary research question was whether
the Medication and CBT treatment groups had better depression outcomes
compared to the TAU group. To answer this question, the data were
analyzed on an intent-to-treat basis using a random intercept and slope
regression model which controlled for ethnicity and baseline
depression. Results from the complete-case analysis showed that both
the Medication intervention ($p<0.001$) and the CBT intervention
($p=0.006$) reduced depression symptoms more than the TAU community referral.

This analysis assumed missing WECare values were MAR. An underlying
concern was whether missing values were nonginorably missing. The
motivation of the work described here was to develop methods of
inference that would reflect uncertainty about the missing data
mechanism in the WECare trial.

\section{Methods}\label{S:methods}
Our approach proceeds in four stages. First, a distribution of
imputation models is specified. Then, nested imputation is conducted in
which $M$ models are drawn from this distribution of models and $N$
multiple imputations for each missing value are generated from each of
the $M$ models resulting in $M \times N$ complete data sets. Next,
parameters of interest are estimated along with their standard errors
for each imputed data set. Finally, the parameter estimates and
standard errors are combined using rules for nested multiple imputation
that yield final inferential results. We also present a method of
quantifying the contribution of missing data mechanism uncertainty to
the overall rate of missing information.

\subsection{Specifying the distribution of imputation models}
The first step in our procedure is identifying a distribution of models
from which it is possible to sample. The choice of which model to use
will depend on subjective notions regarding the dissimilarity of
observed and missing values that the imputer wishes to formalize.
Ideally, this external information is elicited from experts or those
who collected the data.

\citet{rubin87} notes the importance of using easily communicated
models to generate multiple imputations assuming nonignorability so
that users of the completed data can make judgments regarding the
relative merits of the various inferences reached under different
nonresponse models. In this section we describe in detail a method for
generating multiple imputations from multiple models using an
adaptation of a nonignorable imputation procedure suggested
by \citeauthor{rubin87} [(\citeyear{rubin87}), page 22]. In the discussion section we discuss the
application of our multiple model framework using other procedures.

\subsection{Transforming imputed ignorable continuous values to create
nonignorable values}\label{S:constant}

Rubin [(\citeyear{rubin87}), page 203] describes a simple transformation for
generating nonignorable imputed values from ignorable imputed values:
%
\begin{equation}\label{E:constant}
(\mbox{nonignorable imputed } Y_{i})=k \times(\mbox
{ignorable imputed } Y_{i}).
\end{equation}
For example, if $k=1.2$, then the assumption is that, conditioning on
other observed information, missing values are $20\%$ larger than
observed values. In order to create a distribution of nonignorable (and
ignorable) models, we replace the multiplier $k$ in equation (\ref
{E:constant}) with multiple draws from some distribution. If the imputer
believes that missing values tend to be larger than observed values,
then a potential distribution for $k$ might be a $\operatorname{Uniform}(1,3)$
distribution or a $\operatorname{Normal} (1.5,1)$ distribution. By centering the
distribution of $k$ around values smaller than 1.0, nonignorable
imputations can be generated which assume that missing values are
smaller than observed values after conditioning on observed information.

When the ignorable imputed value in equation (\ref{E:constant}) is
negative, the right-hand side of the equation needs to be modified so
that values of $k$ greater than 1 will increase the value of the
ignorable imputed value and values of $k$ less than 1 will decrease the
value of the ignorable imputed value. A more general version of
equation (\ref{E:constant}), applicable in all settings, is
\begin{eqnarray}\label{E:constant_mod}
&&(\mbox{nonignorable imputed } Y_{i})
\nonumber
\\[-8pt]
\\[-8pt]
\nonumber
&&\qquad= [(k-1) \times|\mbox{ignorable imputed }
Y_{i}| ] + \mbox{ignorable imputed } Y_{i}.
\end{eqnarray}
Caution should be exercised to avoid unrealistic imputations.
Multipliers of large magnitude may result in imputations outside the
range of plausible values.

If the imputer wants to generate imputations that are centered around a
missing at random mechanism but with additional uncertainty, they could
specify a $\operatorname{Uniform}(0.5,1.5)$ or $\operatorname{Normal} (1.0,0.25)$ distribution for
the multiplier. More generally, \citet{daniels08} categorize the priors
used in a sensitivity analysis as departures from a MAR mechanism. They
use the following categories: MAR with no uncertainty, MAR with
uncertainty, NMAR with no uncertainty and NMAR with uncertainty. When
viewed in this framework, the standard MAR assumption (MAR with no
uncertainty) is simply one mechanism across a continuum of mechanism
specifications and is equivalent to using a $\operatorname{Normal} (1,0)$ or
$\operatorname{Uniform}(1,1)$ distribution for the multiplier $k$ in equation (\ref
{E:constant_mod}). Note that when we use the term ``imputation model
uncertainty'' we are referring to uncertainty in the missing data
mechanism as governed by uncertainty in the multiplier $k$.

When the data are continuous, equation (\ref{E:constant_mod}) can be
applied to ignorable imputed values that are generated from any
imputation method that assumes ignorability. In this paper we generate
ignorable imputations using regression imputation [\citet{rubin87}, page 166]. We use different values for the multiplier $k$ in
equation~(\ref{E:constant_mod}) to easily generate imputations from many
different models.

\subsection{Nested multiple imputation}\label{S:frame}
Once the distribution of models has been specified, imputation proceeds
in two stages. First $M$ models are drawn from a distribution of models
such as those described in Section \ref{S:constant}. Then $N$ multiple
imputations for each missing value are generated for each of the $M$
models, resulting in $M \times N$ complete data sets.

More specifically, let the complete data be denoted by $Y=(Y_{\mathrm
{obs}},Y_{\mathrm{mis}})$.
For the first stage, the imputation model $\psi$ is drawn from its
predictive distribution
%
\begin{equation}\label{E:model_prior}
\psi^{m} \sim p(\psi),\qquad   m=1, 2, \ldots, M.
\end{equation}

The second stage starts with each model $\psi^{m}$ and draws $n$
independent imputations conditional on $\psi^{m}$,
%
\begin{equation}
Y^{(m,n)}_{\mathrm{mis}} \sim
p(Y_{\mathrm{mis}}|Y_{\mathrm{obs}},\psi^{m}), \qquad
  n=1, 2, \ldots, N.
\end{equation}

Because the $M \times N$ nested multiple imputations are not
independent draws from the same posterior predictive distribution of
$Y_{\mathrm{mis}}$, the traditional multiple imputation combining rules
of Rubin (\citeyear{rubin87}) do not apply. Instead, it is necessary to use combining
rules that take into account variability due to the multiple models.
Fortunately, the method described here is similar to nested multiple
imputation [\citeauthor{harel07} (\citeyear{harel07,harel09b}), \citet{rubin03}, \citet{shen00}]. In the \hyperref[S:nested_just]{Appendix}
we provide further justification for using the nested
imputation combining rules.

\subsection{Combining rules for final inference}\label{S:combine}
In this section we describe the nested multiple imputation combining
rules that we use to combine inferences across multiply imputed data
sets based on multiple imputation models. In describing the rules
below, we use notation that follows closely to that of \citet{shen00}.

Let $Q$ be the quantity of interest. Assume with complete data,
inference about $Q$ would be based on the large sample statement that
\[
(Q-\hat{Q}) \sim N(0,U),
\]
where $\hat{Q}$ is a complete-data statistic estimating $Q$ and $U$ is
a complete-data statistic providing the variance of $Q-\hat{Q}$. The $M
\times N$ imputations are used to construct $M \times N$ completed data
sets, where the estimate and variance of $Q$ from the single imputed
data set is denoted by $(\hat{Q}^{(m,n)},U^{(m,n)})$, where
$m=1,2,\ldots,M$ and $n=1,2,\ldots,N$. The superscript $(m,n)$
represents the $n$th imputed data set under\vadjust{\goodbreak} model $m$. Let $\bar{Q}$ be
the overall average of all $M \times N$ point estimates
%
\begin{equation}\label{E:Qmean}
\bar{Q}=\frac{1}{MN}\sum_{m=1}^{M}\sum_{n=1}^{N}\hat{Q}^{(m,n)},
\end{equation}
and let $\bar{Q}_{m}$ be the average of the $m$th model,
%
\begin{equation}
\bar{Q}_{m}=\frac{1}{N}\sum_{n=1}^{N}\hat{Q}^{(m,n)}.
\end{equation}
Three sources of variability contribute to the uncertainty in $Q$. These
three sources of variability are as follows: $\bar{U}$, the overall
average of the associated variance estimates
%
\begin{equation}\label{E:overall}
\bar{U}=\frac{1}{MN}\sum_{m=1}^{M}\sum_{n=1}^{N}U^{(m,n)},
\end{equation}
$W$, the within-model variance
%
\begin{equation}\label{E:within}
W=\frac{1}{M(N-1)}\sum_{m=1}^{M}\sum_{n=1}^{N}\bigl(\hat{Q}^{(m,n)}-\bar{Q}_{m}\bigr)^{2},
\end{equation}
and $B$, the between-model variance
%
\begin{equation}\label{E:between}
B=\frac{1}{M-1}\sum_{m=1}^{M}(\bar{Q}_{m}-\bar{Q})^{2}.
\end{equation}
The quantity
%
\begin{equation}\label{E:total_var}
T=\bar{U} + \biggl(1+\frac{1}{M}\biggr)B + \biggl(1-\frac{1}{N}\biggr)W
\end{equation}
estimates the total variance of $(Q-\bar{Q})$. Interval estimates and
significance levels for scalar $Q$ are based on a Student-$t$ reference
distribution
\begin{equation}
T^{-{1}/{2}}(Q-\bar{Q})\sim t_{v},
\end{equation}
where $v$, the degrees of freedom, follows from
%
\begin{equation}\label{E:df}
v^{-1}=\biggl[\frac{(1+{1}/{M})B}{T}\biggr]^{2}\frac{1}{M-1}+
\biggl[\frac{(1-{1}/{N})W}{T}\biggr]^{2}\frac{1}{M(N-1)}.
\end{equation}
In standard multiple imputation, only one model is used to generate
imputations so that the between-model variance $B$ [equation (\ref
{E:between})] is equal to 0 and it is not necessary to account for the
extra source of variability due to model uncertainty.

\subsection{Rates of missing information}
Standard multiple imputation provides a rate of missing information
that may be used as a diagnostic measure of how the missing data
contribute to the uncertainty about $Q$, the parameter of\vadjust{\goodbreak}
interest [\citet{schafer97}]. \citeauthor{harel07} (\citeyear{harel07,harel09b}) derived rates of
missing information for nested multiple imputation based on the amount
of missing information due to model uncertainty and missingness. These
rates include an overall rate of missing information $\gamma$, which
can be partitioned into a between-model rate of missing information
$\gamma^{b}$, and a within-model rate of missing information $\gamma
^{w}$. With no missing information (either due to nonresponse or
imputation model uncertainty), the variance of $(Q-\bar{Q})$ reduces to
$\bar{U}$ so that the estimated overall rate of missing information
is [\citet{harel07}]
%
\begin{equation}\label{E:gamma}
\hat{\gamma}=\frac{B + (1-{1}/{N})W}{\bar{U} + B + (1-{1}/{N})W}.
\end{equation}
If the correct imputation model is known, then $B$, the between-model
variance, is~$0$ and the estimated rate of missing information due to
nonresponse is
%
\begin{equation}\label{E:gamma_w}
\hat{\gamma}^{w}=\frac{W}{\bar{U} + W}.
\end{equation}

Roughly speaking, equation (\ref{E:gamma}) measures the fraction of total
variance accounted for by nonresponse and model uncertainty and
equation (\ref{E:gamma_w}) measures the fraction of total variance
accounted for by nonresponse when the correct imputation model is
known. See \citeauthor{harel07} (\citeyear{harel07,harel09b}) for details. The estimated rate of
missing information due to model uncertainty is then $\hat{\gamma
}^{b}=\hat{\gamma}-\hat{\gamma}^{w}$.

In a nested imputation framework, \citet{harel08} takes the ratio $\frac
{\hat{\gamma}^{w}}{\hat{\gamma}}$ which he terms \textit{outfluence}. In
nested imputation, outfluence is a measure of the influence of one type
of missing data relative to all missing values.\vspace*{-1pt} Here, we use the ratio
$\frac{\hat{\gamma}^{b}}{\hat{\gamma}}$ to measure\vspace*{-3pt} the contribution of
model uncertainty to the overall rate of missing information. For
example, a value of $\frac{\hat{\gamma}^{b}}{\hat{\gamma}}$ equal to
0.5 would suggest that half of the overall rate of missing information
is due to missing data mechanism uncertainty, the other half due to
missing values. We anticipate that most researchers would not want to
exceed this value unless they have very little confidence in their
imputation model. Note that most imputation procedures use one model
and implicitly assume that $\frac{\hat{\gamma}^{b}}{\hat{\gamma}}$ is
equal to 0.

In the next section we\vspace*{-1pt} present simulations showing that incorporating
more than one imputation model in an imputation procedure increases
both $\hat{\gamma}^{b}$ and $\frac{\hat{\gamma}^{b}}{\hat{\gamma}}$ and
increases the coverage of parameter estimates versus procedures that
use only one imputation model.

\section{Design of simulation study}\label{S:sim}
In this section we describe a simulation study to illustrate the method
of multiple-model multiple imputation. We simulate longitudinal data
with missing values in order to demonstrate how incorporating missing
data mechanism uncertainty can increase the coverage of parameter estimates.

\subsection{Setup}
Building on an example in \citeauthor{hedeker06} [(\citeyear{hedeker06}), page~283], longitudinal
data with missing values were simulated according to the following
pattern-mixture model:
\begin{eqnarray}\label{E:sim_modl}
y_{ij} &= &\beta_{0} +\beta_{1}\mathrm{Time}_{j} + \beta_{2}\mathrm{Tx}_{i} + \beta
_{3}(\mathrm{Tx}_{i} \times \mathrm{Time}_{j})
\nonumber
\\[-8pt]
\\[-8pt]
\nonumber
&&{} + \beta_{4}(\mathrm{Drop}_{i} \times \mathrm{Time}_{j}) + v_{0i} + v_{1i}\mathrm{Time}_{j} +
\varepsilon_{ij},
\end{eqnarray}
where $\mathrm{Time}_{j}$ was coded 0, 1, 2, 3, 4 for five timepoints, $\mathrm{Tx}_{i}$
was a dummy-coded (i.e., 0 or 1) grouping variable with $150$ subjects
in each group, and $\mathrm{Drop}_{i}$ was a dummy-coded variable indicating
those subjects who eventually dropped out of the study. There were
$100$ dropouts in each treatment group. The regression coefficients
were defined to be as follows: $\beta_{0}=25$, $\beta_{1}=-3$, $\beta
_{2}=0$, $\beta_{3}=-1$, and $\beta_{4}=1.5$. This setup represents a
randomized controlled trial in which group means are equal at baseline
and there is a greater decrease in the outcome measure over time in the
treatment group. Participants who eventually drop out of the study have
smaller decreases in outcomes over time as compared to nondropouts.
Thus, the slope of the treatment and control groups were $-3.0$ and $-2.0$,
respectively. The random subject effects $v_{0i}$ and $v_{1i}$ were
assumed normal with zero means, variances $\sigma^{2}_{v0}=4$ and
$\sigma^{2}_{v1}=1$ and covariance $\sigma_{v01}= -0.1$. The errors
$\varepsilon_{ij}$ were assumed to be normal with mean 0 and variance
$\sigma^{2}=9$ for nondropouts and $\sigma^{2}=16$ for dropouts.

We generated nonignorable missing values on $y_{ij}$ using the
following rule: at timepoints 1, 2, 3 and 4, subjects in the dropout
group dropped out with probabilities (0.25, 0.50, 0.75, 1) so that the
overall proportions of missing values were 0.17, 0.42, 0.60 and 0.67
for the four timepoints. Nondropouts have no missing values at any time
point. The high proportion of dropouts and the use of monotone
missingness (versus intermittent missingness) were chosen so that
post-imputation inferences were sensitive to assumptions regarding the
missing data mechanism.

Imputation using the multiplier approach of Section \ref{S:methods}
proceeded as follows. We first generated 200 imputations of each
missing value using the software package MICE [\citet{MICE}] which
imputes variables one-at-a-time based on a conditional distribution for
each variable. We specified a linear regression model [\citet{rubin87},
page 166] which assumes the missing data are MAR. Each treatment group
was imputed separately to preserve the desirable property in an
intent-to-treat analysis framework that imputed values depend only on
information from other cases in the same treatment arm.

Using the methods described in Sections \ref{S:methods}, we then
transformed the MICE imputations---which assume the data are ignorably
missing---into imputations that assume the data are nonginorably
misssing. Specifically, we simulated 100 values of $k$ from one of the
imputation model distributions listed in Table \ref{Ta:fixed_constant}
and described in Sections \ref{S:ignore_asp} and \ref{S:uncertain_asp}.
Using equation (\ref{E:constant_mod}), each of these values of $k$ was
multiplied to the imputed values in 2 imputed data sets to create 2
imputations nested within 100 models, that is, 200 imputed data sets.

We used $M=100$ imputation models and $N=2$ imputations within each
model so that the degrees of freedom for the within-model variance
$M(N-1)$ [equation~(\ref{E:within})] and the degrees of freedom for the
between-model variance $M-1$ [equation~(\ref{E:between})] were
approximately equal. This allowed us to estimate within- and
between-model variance with equal precision, which is necessary for
stable measurements of the rates of missing information [\citet{harel07}].

We then analyzed the 200 imputed data sets using the random intercept
and slope model described in equation (\ref{E:sim_modl}) but without the
covariates that include dropout. Inferences were combined using the
nested multiple imputation combining rules described in Section \ref
{S:frame}. Here, for brevity, we focus on the slope of the treatment group.

One thousand replications for the above scenario were simulated. An R
function for combining nested multiple imputation inferences and
calculating rates of missing information is available in the
supplementary materials [\citet{siddique12}].

\subsection{Ignorability assumptions}\label{S:ignore_asp}
We explored the effect of imputing under four different ignorability
assumptions which we refer to as MAR, Weak NMAR, Strong NMAR and
Misspecified NMAR. We now discuss each of these assumptions in turn:
\begin{longlist}[(1)]
\item[(1)] Missing at Random (MAR): Under this assumption, we generate
multiple imputations assuming the data are missing at random.
Specifically, we generate imputations assuming the multiplier $k$ in
equation (\ref{E:constant_mod}) is drawn from a distribution with a mean
of 1.0.

\item[(2)] Weak Not Missing at Random (Weak NMAR): Under this assumption, we
generate multiple imputations assuming the data are not missing at
random, but that nonrespondents are not very different from
respondents. Specifically, imputations assuming weak NMAR are generated
by assuming the multiplier $k$ in equation (\ref{E:constant_mod}) is
drawn from a distribution with a mean of 1.3 (nonrespondents have
values that are 30\% larger than respondents).

\item[(3)] Strong NMAR: Here we generate multiple imputations assuming the
data are NMAR and that nonrespondents are quite a bit different than
respondents. Imputations are generated assuming nonrespondents are 70\%
larger than respondents (a multiplier distribution mean of 1.7).

\item[(4)] Misspecified NMAR: Here we generate multiple imputations assuming
the data are NMAR but that nonrespondents have \textit{lower} values than
respondents even though in truth the reverse is true. Imputations
assuming misspecified NMAR are generated by assuming the multiplier $k$
in equation (\ref{E:constant_mod}) is drawn\vadjust{\goodbreak} from a distribution with a
mean of 0.8 (nonrespondents have values that are 20\% smaller than
respondents). We chose this assumption to demonstrate that even when
the imputer is wrong about the nature of nonignorability, incorporating
mechanism uncertainty can increase coverage and make a bad situation better.
\end{longlist}

\subsection{Mechanism uncertainty assumptions}\label{S:uncertain_asp}
In addition to generating imputations using the above ignorability
assumptions, we also generated imputations based on four different
assumptions regarding how certain we were about the correctness of our
models. When there is no mechanism uncertainty, all imputations are
generated from the same model. When there is mechanism uncertainty,
then multiple models are used. All models are centered around one of
the ignorability assumptions in Section \ref{S:ignore_asp}. Uncertainty
is then characterized by departures from the central model. The four
different uncertainty assumptions used to generate multiple models were
as follows: no uncertainty, mild uncertainty, moderate uncertainty and
ample uncertainty. These assumptions are described below:

\begin{longlist}[(1)]
\item[(1)] No uncertainty: This is the assumption of most imputation
schemes. One imputation model is chosen and all imputations are
generated from that one model. In particular, the most common
imputation approach is to assume the data are MAR with no uncertainty.
Imputations with no mechanism uncertainty were generated by using the
same multiplier $k$ in equation (\ref{E:constant_mod}) for all 100
imputation models.

\item[(2)] Mild uncertainty: Here we assume that there is a small degree of
uncertainty regarding what is the right mechanism. By incorporating
some uncertainty into our choice of imputation model, imputations are
generated using multiple models. Specifically, the multiplier $k$ in
equation (\ref{E:constant_mod}) was drawn from a Normal distribution with
a standard deviation of 0.1.

\item[(3)] Moderate uncertainty: Multiple models with moderate uncertainty
are generated using equation (\ref{E:constant_mod}) by drawing the
multiplier from a Normal distribution with a standard deviation of 0.3.

\item[(4)] Ample uncertainty: Multiple models with ample uncertainty are
generated using equation (\ref{E:constant_mod}) by drawing the multiplier
from a Normal distribution with a standard deviation of 0.5.
\end{longlist}

%
\begin{table}
\tabcolsep=0pt
\caption{Simulation study of multiple imputation of continuous data
using multiple models. One hundred models, 2 imputations within each
model}\label{Ta:fixed_constant}
\begin{tabular*}{\textwidth}{@{\extracolsep{\fill}}lccd{2.2}cd{3.1}ccccc@{}}
\hline
\textbf{Ignore} & &\textbf{Model} & &  & &\textbf{Width} &  &  &  &
\\[-4pt]
\textbf{assump.} &\textbf{Uncertainty}&\textbf{Dist'n} & \multicolumn{1}{c}{\textbf{PB}}& \textbf{RMSE}& \multicolumn{1}{c}{\textbf{Cvg.}} &\textbf{of CI} &
$\bolds{\hat{\gamma}}$ &$\bolds{\hat
{\gamma}^{w}}$ & $\bolds{\hat{\gamma}^{b}}$& $\bolds{\frac{\hat{\gamma}^{b}}{\hat
{\gamma}}}$\\
\hline
MAR &None &$N(1.0,0.0)$ & 33.04 & 1.01 & 0.1 & 0.75 & 0.63 & 0.62 & 0.01
& 0.02 \\
&Mild &$N(1.0,0.1)$ & 33.18 & 1.01 & 0.3 & 0.98 & 0.77 & 0.61 & 0.16 &
0.21 \\
&Moderate &$N(1.0,0.3)$ & 33.44 & 1.02 & 53.4 & 2.05 & 0.93 & 0.57 &
0.36 & 0.39 \\
&Ample &$N(1.0,0.5)$ & 33.72 & 1.03 & 99.5 & 3.28 & 0.96 & 0.49 & 0.47 &
0.49 \\[3pt]
Weak &None &$N(1.3,0.0)$ & 18.22 & 0.59 & 36.2 & 0.96 & 0.64 & 0.63 &
0.01 & 0.02 \\
NMAR &Mild &$N(1.3,0.1)$ & 18.35 & 0.59 & 53.5 & 1.14 & 0.74 & 0.62 &
0.12 & 0.16 \\
&Moderate &$N(1.3,0.3)$ & 18.56 & 0.60 & 98.0 & 2.13 & 0.91 & 0.59 &
0.32 & 0.35 \\
&Ample &$N(1.3,0.5)$ & 18.77 & 0.61 & 100.0 & 3.33 & 0.95 & 0.53 & 0.42
& 0.44 \\[3pt]
Strong &None &$N(1.7,0.0)$ & -1.53 & 0.27 & 98.2 & 1.28 & 0.60 & 0.59 &
0.01 & 0.02 \\
NMAR &Mild &$N(1.7,0.1)$ & -1.40 & 0.27 & 99.6 & 1.42 & 0.67 & 0.58 &
0.09 & 0.13 \\
&Moderate &$N(1.7,0.3)$ & -1.19 & 0.27 & 100.0 & 2.29 & 0.86 & 0.56 &
0.29 & 0.34 \\
&Ample &$N(1.7,0.5)$ & -1.03 & 0.28 & 100.0 & 3.42 & 0.92 & 0.53 & 0.40
& 0.43 \\[3pt]
Misspec. &None &$N(0.8,0.0)$ & 42.95 & 1.30 & 0.0 & 0.64 & 0.57 & 0.56 &
0.01 & 0.02 \\
NMAR &Mild &$N(0.8,0.1)$ & 43.10 & 1.30 & 0.0 & 0.90 & 0.77 & 0.56 & 0.22
& 0.28 \\
&Moderate &$N(0.8,0.3)$ & 43.39 & 1.31 & 8.5 & 2.01 & 0.94 & 0.50 & 0.43
& 0.46 \\
&Ample &$N(0.8,0.5)$ & 43.70 & 1.32 & 88.1 & 3.26 & 0.96 & 0.43 & 0.54 &
0.56 \\
\hline
\end{tabular*}
\tabnotetext[]{}{PB: percent bias; RMSE: root mean squared error;
Cvg: coverage.}\vspace*{-3pt}
\end{table}

With four ignorability assumptions and four uncertainty assumptions, we
imputed the data under a total of 16 scenarios. Within each scenario,
we evaluated the percent bias and RMSE of the post-multiple-imputation
treatment slope as well as the coverage rate and width of its nominal
95\% interval estimate. In addition, we calculated measures of missing
information: the overall estimated rate of missing information [$\hat
{\gamma}$ in equation (\ref{E:gamma})], the estimated rate of missing
information due to nonresponse [$\hat{\gamma}^{w}$ in equation (\ref
{E:gamma_w})], the estimated rate of missing information due to model
uncertainty, $\hat{\gamma}^{b}=\hat{\gamma}-\hat{\gamma}^{w}$,\vspace*{-1pt} and the
estimated contribution of model uncertainty to the overall rate of
missing information as measured by the ratio $\frac{\hat{\gamma
}^{b}}{\hat{\gamma}}$.

\section{Simulation results}\label{S:sim_results}
Table \ref{Ta:fixed_constant} lists the results of our imputations
under the 16 different ignorability/uncertainty scenarios using
regression imputation and the methods described in Section \ref
{S:methods} for the slope of the treatment group. Beginning with the
first row, we see that assuming MAR with no mechanism uncertainty
results in estimates that are highly biased with a coverage rate close
to 0\%. This result is not surprising, as the data are nonignorably
missing and here we are assuming in all of our models that the data are
ignorably missing. Since we are using the same model for all
imputations, $\hat{\gamma}^{b}$,\vspace*{-1pt} the estimated fraction of missing
information due to model uncertainty is approximately equal to 0 as is
$\frac{\hat{\gamma}^{b}}{\hat{\gamma}}$, the estimated contribution of
model uncertainty to the overall rate of missing information.

Moving to the subsequent rows in Table \ref{Ta:fixed_constant}, still
assuming MAR, we see the effect of increasing mechanism uncertainty on
post-imputation parameter estimates. Both percent bias and RMSE are the
same as with no uncertainty, but now coverage is increasing as we
increase the amount of uncertainty in our imputation models. Coverage
increases from 0\% to 99.5\%. The mechanism here is clear---by
increasing the amount of uncertainty in our imputation models, we are
now generating imputations under a range of ignorability assumptions.
This additional variability in the imputed values translates to wider
confidence intervals and hence greater coverage. We also see\vspace*{-1pt} that our
measures of missing information are able to pick up this uncertainty.
Both $\hat{\gamma}^{b}$ and $\frac{\hat{\gamma}^{b}}{\hat{\gamma}}$
increase as the amount of model uncertainty increases. As model
uncertainty increases, it becomes a larger proportion of the overall
rate of missing information.

Since missing values in our simulation study tended to be larger than
observed values, the weak and strong NMAR conditions result in smaller
bias than the imputations assuming MAR. As before, increasing the
amount of model uncertainty does not change bias but instead increases
coverage (by increasing the width of the 95\% confidence intervals) to
the point that weak NMAR with moderate and ample uncertainty exceeds
the nominal level. Under the strong NMAR assumption, bias is small
enough that there is no benefit to additional mechanism uncertainty.
Also, as before, additional model uncertainty is reflected in
increasing values of~$\hat{\gamma}^{b}$ and~$\frac{\hat{\gamma
}^{b}}{\hat{\gamma}}$.

Finally, the last four rows of Table \ref{Ta:fixed_constant} present
results when the missing data mechanism is misspecified. Here, the
missing data are imputed assuming that missing values are smaller than
observed values (even after conditioning on observed information) when
in fact the reverse is true. Not surprisingly, bias and RMSE are poor
in this situation, but by incorporating mechanism uncertainly into our
imputations we are able to build some robustness into our imputation
model. With ample uncertainly, coverage is 88.1\%, a substantial
increase over the coverage rate of 0\%, which is the result of using
the same (misspecified) model for all imputations.

%

\section{Application to the Women Entering Care study}\label{S:apply}
We applied our methods to the WECare data as follows. We imputed the
continuous WECare HDRS scores using the same method and imputation
model distribution parameters as described in the simulation study.

The Weak NMAR and Strong NMAR assumptions assume that missing values
tend to be larger than observed values with the same covariates. Since
higher HDRS scores reflect more depression symptoms, these assumptions
imply that nonrespondents are more depressed than respondents even
after conditioning on observed information. The term ``Misspecified''
NMAR is a misnomer in this setting because we do not actually know the
correct specification. We use the term only to be consistent with the
simulation study. For Misspecified NMAR, the assumption is that
nonrespondents are less depressed than respondents.

We investigated how different factors in our imputation procedure
affected inferences from the WECare data. In every scenario, 100 models
were used and~2 imputations were generated within each model for every
missing value. As in the simulation study, each treatment group was
imputed separately.

When imputing and analyzing the WECare data, we restricted our
attention to the depression outcomes that were analyzed in \citet
{miranda03}, variables used as covariates in final analyses, and a set
of additional variables used in the imputation models because they were
judged to be potentially associated with the analysis variables.
Table \ref{Ta:Variables} lists variables that were used in imputation
and analysis models and also indicates the percentage of missing values.

\begin{table}
\caption{WECare variables used for imputation and analysis}\label{Ta:Variables}
\begin{tabular*}{\textwidth}{@{\extracolsep{\fill}}lccc@{}}
\hline
& \textbf{Imputation} & \textbf{Percent} &\textbf{Variable}\\
\textbf{Variable name} & \textbf{or analysis?} & \textbf{missing} &\textbf{type} \\
\hline
Baseline HDRS &Both& \phantom{0}0\% &Scaled \\
Month 1 HDRS &Both& 25\% &Scaled \\
Month 2 HDRS &Both& 24\% &Scaled \\
Month 3 HDRS &Both& 30\% &Scaled \\
Month 4 HDRS &Both& 34\% &Scaled \\
Month 5 HDRS &Both& 38\% &Scaled \\
Month 6 HDRS &Both& 30\% &Scaled \\
Month 8 HDRS & Imputation & 33\% &Scaled \\
Month 10 HDRS & Imputation & 34\% &Scaled \\
Month 12 HDRS & Imputation & 24\% &Scaled \\
Ethnicity& Both &\phantom{0}0\% &Nominal\\
Age & Imputation &\phantom{0}0\% &Continuous\\
Income & Imputation &\phantom{0}4\% &Continuous\\
HS graduate& Imputation &\phantom{0}0\% &Binary \\
Number of children &Imputation &\phantom{0}0\% &Continuous\\
Received 9 wks of Meds &Imputation &\phantom{0}0\% &Binary (Med tx only) \\
No. of CBT sessions &Imputation &\phantom{0}0\% &Continuous (CBT tx only) \\
No. of mental health visits &Imputation &\phantom{0}0\% &Continuous (TAU tx only)
\\
Insurance Status&Imputation&\phantom{0}0\% &Binary\\
Marital Status &Imputation&\phantom{0}0\% &Binary\\
\hline
\end{tabular*}
\tabnotetext[]{}{HDRS: Hamilton depression rating scale.}
\end{table}

Four important targets of inference from the random intercept and slope
model used in \citet{miranda03} are the slopes of the Medication
treatment group and the CBT treatment group, reflecting the change in
HDRS scores over time for the two active interventions and their
difference with the slope of the TAU condition, which estimates the
effect of treatment. Here, for brevity, we focus our attention on the
slope of the Medication treatment group and also its difference with
the slope of the TAU group (i.e., the Medication treatment effect) to
illustrate the impact of different ignorability and uncertainty
assumptions in our imputation procedures.

\subsection{Imputation of HDRS scores}
Imputation of the monthly HDRS scores using the multiplier approach of
Section \ref{S:methods} proceeded as follows. For every
ignorabilty/uncertainty combination in Table \ref{Ta:fixed_constant},
we first generated 200 imputations of the WECare missing data using
MICE [\citet{MICE}] and specified a linear regression model [\citet
{rubin87}, page~166] to impute income and depression scores. This
method assumes the missing data are MAR. Each imputation model
conditioned on all the variables listed in Table \ref{Ta:Variables}. In
particular, depression scores were imputed using a model that
conditioned on both prior depression scores and subsequent depression
scores in order to make use of all available information. Imputed
values were rounded to the nearest observed value to create plausible
HDRS scores.

We then simulated 100 values from the corresponding
ignorability/uncer\-tainty distributions listed in Table \ref
{Ta:fixed_constant} and described in Sections \ref{S:ignore_asp}
and \ref{S:uncertain_asp}. Using equation~(\ref{E:constant_mod}), each of
these values of $k$ was multiplied to the imputed values in 2 imputed
data sets to create 2 imputations nested within 100 models. Many of the
ignorability/uncertainty distributions that are used in the simulation
are not realistic for this application, but we use them here for the
sake of brevity and so that we can clearly see the effect of different
assumptions on post-imputation inferences. Imputed values were again
rounded to the nearest observed value to create plausible HDRS scores.
We then analyzed the 200 imputed data sets using the random intercept
and slope regression model of \citet{miranda03}, and the nested
imputation combining rules described in Section \ref{S:combine}.

\begin{table}
\tabcolsep=4pt
\caption{Post-imputation WECare Medication intervention slopes by
ignorability/uncertainty scenario. One-hundred models with 2
imputations per model were used to generate 200 imputations.
Multipliers were generated by drawing from a Normal distribution. MAR,
Weak NMAR, Strong NMAR and Misspecified NMAR correspond to Normal
distributions with means of 1, 1.3, 1.7 and 0.8, respectively. Amounts
of uncertainty None, Mild, Moderate, Ample correspond to Normal
distributions with standard deviations of 0, 0.1, 0.3 and 0.5,
respectively}\label{Ta:wecare_mult_meds} 
\begin{tabular*}{\textwidth}{@{\extracolsep{\fill}}lcd{2.2}cd{2.2}d{2.2}d{2.2}cccc@{}}
\hline
\textbf{Ignore}& & & & & & & & & &\\[-4pt]
\textbf{assump.} &\textbf{Uncertainty} &\multicolumn{1}{c}{\textbf{Est.}}&\textbf{SE} & \multicolumn{1}{c}{\textbf{LCI}} &\multicolumn{1}{c}{\textbf{UCI}} &
\multicolumn{1}{c}{$\bolds{p}$\textbf{-val.}} & $\bolds{\hat{\gamma}}$ & $\bolds{\hat{\gamma
}^{w}}$ & $\bolds{\hat{\gamma}^{b}}$ & $\bolds{\frac{\hat{\gamma}^{b}}{\hat{\gamma}}}$ \\
\hline
MAR & None &-1.93 & 0.47 & -2.86 & -1.00 & <\!0.01 & 0.37 & 0.36 & 0.01
& 0.03 \\
& Mild &-1.95 & 0.53 & -3.00 & -0.91 & <\!0.01 & 0.49 & 0.37 & 0.13 &
0.25 \\
& Moderate &-2.02 & 0.85 & -3.70 & -0.35 & 0.02 & 0.77 & 0.35 & 0.42 &
0.54 \\
& Ample &-2.09 & 1.20 & -4.46 & 0.28 & 0.08 & 0.87 & 0.32 & 0.54 &
0.63 \\[3pt]
Weak & None &-1.71 & 0.56 & -2.81 & -0.61 & <\!0.01 & 0.42 & 0.41 &
0.01 & 0.03 \\
NMAR & Mild &-1.74 & 0.60 & -2.91 & -0.57 & <\!0.01 & 0.49 & 0.41 &
0.08 & 0.16 \\
& Moderate &-1.82 & 0.84 & -3.46 & -0.17 & 0.03 & 0.72 & 0.40 & 0.33 &
0.45 \\
& Ample &-1.91 & 1.15 & -4.17 & 0.35 & 0.10 & 0.84 & 0.36 & 0.47 & 0.57
\\[3pt]
Strong & None &-1.53 & 0.65 & -2.80 & -0.25 & 0.02 & 0.42 & 0.40 & 0.01
& 0.03 \\
NMAR & Mild &-1.54 & 0.66 & -2.84 & -0.24 & 0.02 & 0.45 & 0.40 & 0.04 &
0.09 \\
& Moderate &-1.61 & 0.80 & -3.19 & -0.03 & 0.05 & 0.62 & 0.40 & 0.22 &
0.35 \\
& Ample &-1.70 & 1.04 & -3.74 & 0.34 & 0.10 & 0.76 & 0.38 & 0.39 & 0.51
\\[3pt]
Misspec. & None &-2.10 & 0.42 & -2.93 & -1.27 & <\!0.01 & 0.30 & 0.29 &
0.01 & 0.03 \\
NMAR & Mild &-2.12 & 0.49 & -3.09 & -1.16 & <\!0.01 & 0.47 & 0.30 &
0.17 & 0.37 \\
& Moderate &-2.18 & 0.85 & -3.85 & -0.51 & 0.01 & 0.79 & 0.29 & 0.49 &
0.63 \\
& Ample &-2.22 & 1.20 & -4.59 & 0.16 & 0.07 & 0.87 & 0.28 & 0.59 & 0.68
\\
\hline
\end{tabular*}
\tabnotetext[]{}{SE: standard error; LCI: lower 95\% confidence
interval; UCI: upper 95\% confidence interval.}
\end{table}

\subsection{Post multiple imputation results from the WECare analysis}
Table \ref{Ta:wecare_mult_meds} provides estimates, standard errors,
confidence intervals, $p$-values and rates of missing information for the
WECare Medication slope by the 16 different ignorability/uncertainty
scenarios described in Sections \ref{S:ignore_asp} and \ref
{S:uncertain_asp} using the multiple model approach described in
Section \ref{S:methods}. Table \ref{Ta:wecare_mult_medtx} provides the
same information for the difference between the Medication and TAU slopes.

Looking first at Table \ref{Ta:wecare_mult_meds}, we see that
assumptions regarding ignorability and uncertainty have an impact on
parameter estimates and their associated standard errors. Starting with
those rows assuming MAR, we see that the point estimate for the slope
changes very little for all four uncertainty assumptions. However, as
we assume more uncertainty, the associated standard errors increase.
This same phenomenon was seen in the simulation study. The additional\vspace*{-1pt}
model uncertainty is also reflected in increasing values of $\hat{\gamma
}^{b}$ and $\frac{\hat{\gamma}^{b}}{\hat{\gamma}}$, the estimated rate
of missing information due to model uncertainty and the estimated
contribution of model uncertainty to the overall rate of missing
information, respectively. These values are quite large under ample
uncertainty, reflecting the fact that the ample uncertainty assumption
is relatively diffuse for these data. Because of this, for every
ignorability scenario, ample uncertainty results in slopes that are no
longer significantly different from 0 at the 0.05 level.

As mentioned above, the Weak NMAR and Strong NMAR assumptions assume
that nonrespondents are more depressed than respondents even after
conditioning on observed information. Since there are more missing
values later in the study, these assumptions have the effect of
flattening the slope of the Medication intervention. Within any
ignorability assumption, the point estimates of the slope change only a
little but standard errors increase as more model uncertainty is
assumed. Again, the values of $\hat{\gamma}^{b}$ and $\frac{\hat{\gamma
}^{b}}{\hat{\gamma}}$ appear to capture this uncertainty.

The ``Misspecified'' NMAR assumption assumes that nonrespondents are
less depressed than respondents and, as a result, the slope estimate is
steeper than any of the other scenarios.

\begin{table}
\tabcolsep=4pt
\caption{Post-imputation WECare Medication intervention treatment
effects by ignorability/uncertainty scenario. One hundred models with 2
imputations per model were used to generate 200 imputations.
Multipliers were generated by drawing from a Normal distribution. MAR,
Weak NMAR, Strong NMAR and Misspecified NMAR correspond to Normal
distributions with means of 1, 1.3, 1.7 and 0.8, respectively. Amounts
of uncertainty None, Mild, Moderate, Ample correspond to Normal
distributions with standard deviations of 0, 0.1, 0.3 and 0.5,
respectively}\label{Ta:wecare_mult_medtx}
\begin{tabular*}{\textwidth}{@{\extracolsep{\fill}}lcd{2.2}cd{2.2}d{2.2}d{2.2}cccc@{}}
\hline
\textbf{Ignore}& && & & & & & & & \\[-4pt]
\textbf{assump.} &\multicolumn{1}{c}{\textbf{Uncertainty}} &\multicolumn{1}{c}{\textbf{Est.}} &\multicolumn{1}{c}{\textbf{SE}} &
\multicolumn{1}{c}{\textbf{LCI}} &\multicolumn{1}{c}{\textbf{UCI}} & \multicolumn{1}{c}{$\bolds{p}$\textbf{-val.}} &
$\bolds{\hat{\gamma}}$ & $\bolds{\hat{\gamma
}^{w}}$ & $\bolds{\hat{\gamma}^{b}}$ & $\bolds{\frac{\hat{\gamma}^{b}}{\hat{\gamma}}}$ \\
\hline
MAR & None &-0.69 & 0.25 & -1.18 & -0.19 & <\!0.01 & 0.34 & 0.34 & 0.00
& 0.00 \\
& Mild &-0.69 & 0.27 & -1.22 & -0.17 & 0.01 & 0.42 & 0.35 & 0.06 &
0.15 \\
& Moderate &-0.70 & 0.38 & -1.46 & 0.05 & 0.07 & 0.69 & 0.34 & 0.35 &
0.51 \\
& Ample &-0.71 & 0.52 & -1.73 & 0.31 & 0.17 & 0.81 & 0.31 & 0.49 &
0.61 \\[3pt]
Weak & None &-0.70 & 0.30 & -1.29 & -0.11 & 0.02 & 0.37 & 0.37 & 0.00 &
0.00 \\
NMAR & Mild &-0.71 & 0.31 & -1.31 & -0.10 & 0.02 & 0.41 & 0.38 & 0.02 &
0.05 \\
& Moderate &-0.71 & 0.39 & -1.48 & 0.05 & 0.07 & 0.62 & 0.37 & 0.25 &
0.40 \\
& Ample &-0.72 & 0.51 & -1.72 & 0.29 & 0.16 & 0.77 & 0.35 & 0.41 & 0.54
\\[3pt]
Strong & None &-0.70 & 0.35 & -1.39 & -0.00 & 0.05 & 0.36 & 0.36 & 0.00
& 0.00 \\
NMAR & Mild &-0.70 & 0.35 & -1.39 & -0.01 & 0.05 & 0.37 & 0.37 & 0.00 &
0.00 \\
& Moderate &-0.71 & 0.40 & -1.49 & 0.07 & 0.07 & 0.50 & 0.38 & 0.12 &
0.24 \\
& Ample &-0.71 & 0.48 & -1.66 & 0.23 & 0.14 & 0.66 & 0.37 & 0.29 & 0.44
\\[3pt]
Misspec. & None &-0.67 & 0.22 & -1.12 & -0.23 & <\!0.01 & 0.27 & 0.27 &
0.00 & 0.00 \\
NMAR & Mild &-0.68 & 0.25 & -1.16 & -0.20 & <\!0.01 & 0.38 & 0.29 &
0.10 & 0.26 \\
& Moderate &-0.69 & 0.38 & -1.43 & 0.05 & 0.07 & 0.71 & 0.29 & 0.42 &
0.59 \\
& Ample &-0.70 & 0.52 & -1.72 & 0.32 & 0.18 & 0.82 & 0.27 & 0.55 & 0.67
\\
\hline
\end{tabular*}
\tabnotetext[]{}{SE: standard error; LCI: lower 95\% confidence
interval; UCI: upper 95\% confidence interval.}
\end{table}

Table \ref{Ta:wecare_mult_medtx} displays results for the difference
between the Medication and TAU slopes. For this quantity, the point
estimate is almost the same in every ignorability/uncertainty scenario.
This result is not surprising, as there were similar amounts of missing
Medication and TAU data at each timepoint. For each ignorability
assumption, the slope of the TAU intervention changed by the same
magnitude as the slope of the Medication intervention. As a result,
their difference remains constant at each assumption. However,
incorporating model uncertainty into the imputations does increase the
standard error of this parameter estimate. In fact, under moderate and
ample uncertainty the treatment effect of the Medication intervention
is no longer significant at the $0.05$ level. These results underscore
the importance of making reasonable assumptions. As noted above, the
uncertainty assumptions in this example were chosen to be consistent
with the simulation study and may not be realistic in a depression study.

In the scenarios in Table \ref{Ta:wecare_mult_medtx} where there was no
model uncertainty, the original estimates of the rate of missing
information due to model uncertainty were negative. As noted by \citet
{harel09}, this is possible due to the use of the method of moments for
calculating the rates of missing information.\vspace*{-1pt} Following their
recommendation, we set $\hat{\gamma}^{b}$ and $\frac{\hat{\gamma
}^{b}}{\hat{\gamma}}$ equal to 0 when $\hat{\gamma}^{b}$ was
negative.\eject

%

\section{Discussion}\label{S:discuss}
We have described a relatively simple method for generating multiple
imputations in the presence of nonignorable nonresponse. By generating
multiple imputations from multiple models, our method allows the user
to incorporate uncertainty regarding the missing data mechanism into
their parameter estimates. This is a useful approach when the missing
data mechanism is unknown, which is almost always the case with
nonignorably missing data. Our goal was not to develop a competitor to
model-based methods such as selection models and pattern-mixture
models. Instead, we wished to provide a imputation-based alternative to
model-based methods for those researchers who prefer to use
complete-data methods.

As seen in both the simulation studies and the application to the
WECare data, post-imputation inferences can be highly sensitive to the
choice of the imputation model. With the WECare data, imputation using
our methods had a strong effect on the slope of the Medication
intervention but little effect on the difference in slopes between the
Medication and TAU groups. However, the Medication treatment effect was
no longer significant when moderate and ample imputation model
uncertainty were assumed.

This ability to render nonsignificant a result that is significant
assuming ignorability (and vice versa) suggests that careful attention
should be paid to the specification of the imputation model in
equation (\ref{E:model_prior}). It may make sense to have analysis
protocols specify clearly in advance what missing data assumptions will
be explored. Imputation model assumptions should be chosen prior to
analysis and not based on whether it produces the desired result. Here,
the literature on prior elicitation may be helpful [\citet
{kadane98}, \citet{paddock09}, \citet{white07}].

One approach for eliciting expert opinion when choosing a distribution
for the multiplier $k$ in equation (\ref{E:constant_mod}) is to ask a
subject-matter expert to provide an upper and lower bound for the
multiplier. Then, assuming the multiplier is normally distributed, set
the multiplier distribution mean equal to the average of the lower and
upper bounds, and the standard deviation equal to the difference in
bounds divided by 4. This assumes that the range defined by the upper
and lower bounds is a 95\% confidence interval which may be appropriate
given the tendency of people to specify overly narrow confidence
intervals [\citet{tversky74}]. A~similar calculation can be used if
assuming a uniform prior.

Once the data have been imputed, it is important to examine rates of
missing information, in particular, $\hat{\gamma}^{b}$ and $\frac{\hat
{\gamma}^{b}}{\hat{\gamma}}$, to confirm that appropriate uncertainty
is being incorporated into imputations. For example, if imputations
outside the range of possible values are rounded up or down to the
nearest observed value, this could result in too little variability,
resulting in decreased coverage.

One approach for ensuring that appropriate uncertainty is incorporated
into inferences is to generate imputations and perform analyses based
on a few different distributions for the multiplier. Then, without
examining the significance of parameter estimates, confirm that
appropriate imputation model uncertainty is being incorporated into the
parameter estimates. Because our methods begin with the same set of
ignorable imputations, it is relatively easy to generate imputations
using different missing data mechanisms.

Our approach uses a large number of imputation models $M$, as this is
necessary to obtain stable estimates of the rates of missing
information. The relative (compared to an infinite number of
imputations) efficiency of point estimates using nested multiple
imputation is a function of the fraction of missing information as well
as $M$ and $N$. Improvements in relative efficiency are minimal when
one uses more than a modest number of imputations. Hence, when the
researcher's main interest is point estimates and their variances, a
smaller number of imputations are usually sufficient, for example,
$M=10\mbox{--}20$ and $N=2$ [\citet{harel07}].

In line with more of a sensitivity analysis rather than a final
analysis, when it is hard to pin down a single range for the
multiplier, one may consider a growing set of ranges and observe how
subsequent inferences evolve accordingly. This approach will allow the
user to make more precise statements regarding the exact conditions
under which the obtained results apply [\citet{vanbuuren99}].

Although we believe that all imputation model uncertainty should be
incorporated into one inference, our approach is not inconsistent with
a sensitivity analysis that examines inferences across a range of
ignorability assumptions. \citet{scharfstein99} view sensitivity
analysis as useful ``preprocessing'' for any full Bayesian analysis that
places prior distributions on sensitivity parameters and recommend that
one also publish the results based on the individual sensitivity
parameters in addition to the results that average across a range of
sensitivity parameters so that readers are aware of how inferences vary
based on individual sensitivity parameters.

Our approach is less extreme than worst-case best-case intervals [\citet{cochran77}, page 361]
because we allow for imputation model parameters
to fall within a chosen range in order to obtain narrower and more
plausible ranges of estimates. Including implausible imputation model
parameters broadens the range of inferences unnecessarily and can
introduce implausible values. Instead, our imputation models are given
appropriate weight, with imputation models that lead to extreme
scenarios receiving less weight than models that lead to less extreme
alternatives.

Of course, in any applied setting it is impossible to know exactly how
strong a nonignorable assumption one should make and how much
uncertainty one should place on their models. We see the second of
these dilemmas---incorporating appropriate mechanism uncertainty---as
deserving more attention. Attempting to correctly specify the missing
data mechanism is difficult in most settings. Still, we see our method
as an improvement over methods that make no assumptions regarding
missing data mechanism uncertainty. In addition, our method provides
easily stated subjective notions regarding nonresponse so that they can
be easily stated, communicated and compared.

We see a number of possible variations of our approach. For example, in
some longitudinal data settings, it may be appropriate to use ignorable
models early in the study, and nonignorable models later in the study,
or perhaps incorporate less mechanism uncertainty early in the study
and more later in the study.

Another possible approach is to use different imputation models for
different groups of participants. For example, in the WECare study, we
might want to generate nonignorable imputations for dropouts and
ignorable imputations for everyone else. If the reasons for missingness
are thought to differ by treatment group, it may be appropriate to use
different assumptions for each treatment group. If one believes that
nonresponse is due to both NMAR and MAR mechanisms [\citet{barnes10}],
one could draw the multiplier from a mixture of distributions centered
around both MAR and NMAR assumptions.

When an analyst has prior beliefs about the nature of missingness at a
given time point given what occurred at previous time points, careful
thought should go into the choice of the imputation model and
multiplier distribution. Uncertainty regarding these beliefs can also
be incorporated into the multiple models framework. Alternatively,
methods that explicitly model this temporal relationship such as
selection models and pattern-mixture models may be more
appropriate [\citet{molenberghs03}, \citet{thijs02}].

Some other approaches for generating multiple-model multiple
imputations that can be incorporated into our framework include mixture
model imputation [\citet{rubin87}, \citet{vanbuuren99}], imputation based on a
multivariate $t$-distribution with varying degrees of freedom [\citet
{liu95}] and pattern-mixture model imputation [\citet{demirtas03}, \citet{thijs02}]. \citet{carpenter07} propose an extension to
their method where the multiple reweighting parameters are drawn from a
Normal distribution to incorporate uncertainty in the sensitivity
parameter. Finally, a nonignorable approximate Bayesian bootstrap [\citet
{rubin91}, \citet{siddique08b}] in conjunction with hot-deck imputation can be
also be used. This approach has the added benefit of generating
plausible imputed values since imputations are based on values observed
elsewhere. An important consideration when developing methods for
generating nonignorable imputations is that as the methods become more
complex, it becomes harder to communicate exactly how imputations were
generated and the payoff for the additional complexity is not always clear.

\begin{appendix}
\section*{Appendix: Motivation for using nested multiple imputation}\label{S:nested_just}
In this section we provide motivation for using the nested multiple
imputation combining rules. As in Section \ref{S:methods}, let $Q$ be
the quantity of interest, $Y_{\mathrm{mis}}$ represent the missing values and
$\psi$ the imputation model. The observed data posterior of $Q$ using
our approach is
%
\begin{eqnarray}\label{E:posterior}
p(Q|Y_{\mathrm{obs}}) &= &\int\int p(Q|Y_{\mathrm{obs}},Y_{\mathrm{mis}},\psi) p(Y_{\mathrm{mis}},\psi
|Y_{\mathrm{obs}})\,dY_{\mathrm{mis}}\,d\psi
\nonumber
\\[-8pt]
\\[-8pt]
\nonumber
&= &\int\int p(Q|Y_{\mathrm{obs}},Y_{\mathrm{mis}},\psi) p(Y_{\mathrm{mis}}|Y_{\mathrm{obs}},\psi)p(\psi
)\,dY_{\mathrm{mis}}\,d\psi.
\end{eqnarray}
Note the posterior distribution of $Y_{\mathrm{mis}}$, $p(Y_{\mathrm{mis}}|\psi
,Y_{\mathrm{obs}})$, conditions on $\psi$ so that nested multiple imputations
are not independent draws from the same posterior distribution. When
the posterior mean and variance are adequate summaries of the posterior
distribution, equation (\ref{E:posterior}) can be effectively replaced by
%
\begin{equation}\label{E:post_mean}
E(Q|Y_{\mathrm{obs}}) = E(E(E(Q|Y_{\mathrm{obs}},Y_{\mathrm{mis}},\psi)|Y_{\mathrm{obs}},\psi))
\end{equation}
and
%
\begin{eqnarray}
\operatorname{Var}(Q|Y_{\mathrm{obs}}) &=& E(\operatorname{Var}(Q|Y_{\mathrm{obs}},Y_{\mathrm{mis}},\psi)) +
\operatorname{Var}(E(Q|Y_{\mathrm{obs}},Y_{\mathrm{mis}},\psi))
\nonumber
\\
&= &E(E(\operatorname{Var}(Q|Y_{\mathrm{obs}},Y_{\mathrm{mis}},\psi)|Y_{\mathrm{obs}},\psi)) \label{E:comp_var} \\
&&{}+ E(\operatorname{Var}(E(Q|Y_{\mathrm{obs}},Y_{\mathrm{mis}},\psi)|Y_{\mathrm{obs}},\psi)) \label{E:with_var} \\
&&{}+ \operatorname{Var}(E(E(Q|Y_{\mathrm{obs}},Y_{\mathrm{mis}},\psi)|Y_{\mathrm{obs}},\psi)) \label{E:betw_var}.
\end{eqnarray}
The three variance components in equations (\ref{E:comp_var}), (\ref{E:with_var}) and (\ref{E:betw_var}) correspond to the the overall average
complete data variance, the within-model variance and the between-model
variance, respectively.

The mean in equation (\ref{E:post_mean}) is approximated using
equation (\ref{E:Qmean}). And the variance components in equations (\ref
{E:comp_var}), (\ref{E:with_var}) and (\ref{E:betw_var}) are approximated
using equations (\ref{E:overall}), (\ref{E:within}) and (\ref{E:between}) in
Section \ref{S:combine}.
\end{appendix}

\section*{Acknowledgments}
The authors wish to thank the Editor and two anonymous reviewers whose
comments greatly improved the quality of the manu\-script. The authors
also wish to thank Jeanne Miranda for use of the WECare data.

\begin{supplement}[id=suppA]
\stitle{CombineNestedImputations: An R function for combining
inferences based on nested multiple imputations}
\slink[doi]{10.1214/12-AOAS555SUPP} 
\slink[url]{http://lib.stat.cmu.edu/aoas/555/CombineNestedImputations.R}
\sdatatype{.R}
\sdescription{This R function combines inferences based on nested
multiply imputed data sets and calculates rates of missing information.}
\end{supplement}

%


\printaddresses

\end{document}